\documentclass[useAMS,usenatbib]{mn2e}

\usepackage{amssymb,amsmath,upgreek,bm,color}

\usepackage{epsfig,graphicx,times,subfigure}

\newcommand{\aaps}{A\&A}
\newcommand{\aj}{AJ}
\newcommand{\pasp}{PASP}
\newcommand{\apj}{ApJ}
\newcommand{\mnras}{MNRAS}
\newcommand{\araa}{ARA\&A}
\newcommand{\apjs}{ApJS}
\newcommand{\nat}{Nature}

\usepackage{gensymb}

\title[An ETG with an inner star-forming disk]
{An early-type galaxy with an inner star-forming disk}

\author[Li S.~L. et al.]{
Song-lin Li,$^{1,2}$
Yong Shi,$^{1,2}$\thanks{E-mail:yong@nju.edu.cn}
Yan-Mei Chen,$^{1,2}$\thanks{E-mail:chenym@nju.edu.cn}
Martha Tabor,$^{3}$
Dmitry Bizyaev,$^{4,5,6}$
\newauthor
Jian-hang Chen,$^{1,2}$
Xiao-ling Yu,$^{1,2}$
Long-ji Bing$^{1,2}$ \\
$^{1}$School of Astronomy and Space Science, Nanjing University, Nanjing 210093, China.\\
$^{2}$Key Laboratory of Modern Astronomy and Astrophysics (Nanjing University), Ministry of Education, Nanjing 210093, China.\\
$^{3}$School of Physics and Astronomy, University of Nottingham, University Park, Nottingham, NG7 2RD, UK.\\
$^{4}$Apache Point Observatory and New Mexico State University, P.O. Box 59, Sunspot, NM, 88349-0059, USA.\\
$^{5}$Sternberg Astronomical Institute, Moscow State University, 119234, Moscow, Russia.\\
$^{6}$Special Astrophysical Observatory of the Russian AS, 369167, Nizhnij Arkhyz, Russia.\\
	}

\begin{document}

\date{Submitted to Monthly Notices of the Royal Astronomical Society}

\pagerange{\pageref{firstpage}--\pageref{lastpage}} \pubyear{}

\maketitle
\newpage
\label{firstpage}
\newpage
\pagebreak
\begin{abstract}

  Early-type galaxies (ETGs)  are composed of two distinct  populations: high-mass and low-mass, which are likely to be built via gas-poor merging
  and  gas-rich   merging/accretion, respectively.  However,   it  is
  difficult to  directly associate low-mass ETGs with gas-rich
  processes,  because currently they are gas  poor  with no signs of ongoing star
 formation. We report a  discovery of an ETG (SDSS J142055.01+400715.7) with $M_{*}$=10$^{10}$
  M$_{\odot}$ that  offers direct  evidence for gas-rich merging  as the
  origin  of low-mass  ETGs. The  integrated properties  of the
  galaxy are  consistent with a  typical low-mass ETG, but the
  outer and inner regions show distinct dispersion- and
  rotation-dominated kinematics, respectively. There are some tidal features surrounding the galaxy. These two facts suggest very recent galaxy
  merging. Furthermore, the inner disk harbors on-going star formation,
  indicating the  merging to be  gas rich. This  type of  galaxy is
  rare but it may be a demonstration of the role the transient phase of gas-rich merging plays in making a
  low-mass ETG.

\end{abstract}


\section{Introduction}

The   Hubble    sequence   of   galaxies   as    first   proposed   by
\cite{Hub:26,Hub:36}  divides  galaxies  into two  types:  late-type
 and early-type galaxies (ETGs), characterised by whether or not they contain spiral
features. ETGs   are traditionally  classified into
ellipticals and S0s depending on the  presence of a disk, but the
photometric analysis  of galaxies' isophotes suggests  that some ellipticals
can  still  contain weak  disk  components  \citep{Ben:89}, while
inclination  effects  can lead to the mis-classification  of some  S0s   as  ellipticals
\citep{Jor:94}.

Among ETGs, high-mass ($M_{*}$  $>$ 10$^{11}$ M$_{\odot}$, $M_{\rm V}$
$\lesssim$  -21.5) and  low-mass galaxies are  found to  be two  distinct
galaxy  populations.   Massive  ETGs  always have  boxy  and  triaxial
isophotal shapes while low-mass  ones are disky \citep{Kor:99, Kor:09}
and axisymmetric \citep{Cap:13a}.   In terms of the  radial profile of
the optical light, massive ETGs have relatively large S\'ersic indices
at the outer radius  \citep[$\rm n\geq4$][]{Kra:13a} but show a deficit
at the inner  radius, known as the core.  On  the other hand,
low-mass  ETGs show  relatively small  S\'ersic indices  at the  outer
radius ($\rm n\leq3$ \citet{Kra:13a}), and  are core-less with either a
power-law inner profile or even extra nuclear light \citep{Kor:99}.
Observations  further indicate  that some  central components  have
younger   stellar  populations   than   the  rest   of  the   galaxies
\citep{Lau:05,   Mcd:06}. 

With   integral   field  unit   (IFU)
spectroscopic observations  such as  the SAURON survey  \citep{deZ:02} and
ATLAS$^{\rm 3D}$  project \citep{Cap:11a}, it is  possible to classify
ETGs  according  to  their  2D  kinematic  properties.   For  example,
\cite{Kra:06,Kra:11} found  that ETGs can be  separated into ``regular
rotators'' with a regular  velocity field and ``non-regular rotators''
dominated by random motion,  associated with  high-mass and low-mass
ETGs, respectively.  \citet{Ems:07} used  the apparent angular momentum parameter
to  separate ETGs  and found  that  massive ETGs  rotate slowly  (slow
rotators)  and low-mass  ellipticals  rotate  faster (fast  rotators).
\cite{Cap:11a,Cap:11b}  argued  that  fast   rotators  form  a  smooth
parallel sequence to  spiral galaxies on the  luminosity/mass vs. size
plane.

It  is  suspected   that  the  cores  of  massive   ETGs  form  through
dissipationless  dry mergers  in which  supermassive black  holes sink
into  the  galaxy center  to   eject  stars  and   form  a   nuclear  core
\citep{Kor:09,  Cap:16, Kra:13b},  while  extra  nuclear light  in
low-mass ETGs is produced by gas-rich merging or accretion, in which a
nuclear starburst is  triggered.  However, all ETGs,  remnants of both
gas-poor and  gas-rich merging,  lack on-going  star formation
since even in the case of gas-rich merging the gas has been dispelled or consumed, ceasing star formation.  
This makes  it difficult  to   directly  associate  a  low-mass   ETG  with  gas-rich
processes. In the  MaNGA project (Mapping Nearby Galaxies  at Apache Point
Observatory ),   we    identified   an   ETG  SDSS J142055.01+400715.7
({\it RA=14h20m55.014s \& DEC=+40d07m15.70s}), which may be a living example
that gas-rich merging is playing a major role in forming a low-mass ETG with extra nuclear light.

In  section  2  we  briefly   introduce  the  MaNGA  project, the
characteristics  of the  galaxy we  focus  on and the methodology. In  section 3  we list  the
physical  properties  of  this  galaxy. Followed  the  discussion  in
section 4, the summary and conclusion is in section 5.
 
\section{Data and Methodology}
\label{sect:manga}

MaNGA \citep{Bun:15, Law:15} is a part of Sloan Digital Sky Survey  IV (SDSS-IV) \citep{Bla:17}. This survey has started in July of 2014 on the 2.5m telescope at Apa he Point Observatory (APO) \citep{Gun:06}. Its aim is to investigate
the internal kinematic  structure and composition of gas  and stars in
an  unprecedented sample  of 10,000  nearby galaxies  in the  redshift
range  0.01$<$z$<$0.15  through  IFU spectroscopic  observations \citep{Wak:17}.  The
MaNGA instrument consists a set  of 17 hexagonal fiber bundle integral
field units  that vary in diameter  from 12 arcsec (19 fibers)  to 32 arcsec
(127 fibers), with 12  7-fiber ``mini-bundles'' for spectrophotometric
calibration and 92  single fibers for sky subtraction \citep{Dro:15,Yan:16a}.  All fibers are
fed  into the dual beam BOSS spectrographs,  covering a  wavelength
range  from  3600\AA  $\;$to  10300\AA  $\;$ with  spectral  resolution
R$\sim$2000 \citep{Sme:13}. The raw data were reduced though data reduction pipeline described in \cite{Law:16}. 

The SDSS J142055.01+400715.7 is located  at {\it z=0.01754}, being a member of the primary sample with the coverage out to 1.5 $\rm R_e$ \citep{Yan:16b}.   The  FWHM  of  the reconstructed PSF of our data  cube is $\sim$2.4 arcsec, corresponding to
a physical scale  of 0.86 kpc. We
have performed the following measurements:

{\bf IRAF Ellipse fitting:} We  used the ``Ellipse'' procedure in IRAF\footnote{The IRAF homepage is here: http://iraf.noao.edu.}
to obtain  the 1-D  profiles of the  galaxy in $g$,  $r$ and $i$ 
bands. The contamination
of companion galaxies  and foreground stars in the  field were removed
automatically.  We then fitted the  profiles with two S\'ersic models
\citep{Ser:68}. To take the point spread function (PSF) effect into account and get the intrinsic S\'ersic indices, we simplified the PSF as a single Gaussian function and convolved it with two S\'ersic models, then used this convolved model to fit these 1-D profiles.

{\bf $\bm{\lambda-\epsilon}$ diagram: } 
\cite{Ems:07} proposed a simple parameter $\rm \lambda_R$ to trace the stellar angular momentum of a galaxy.
For the IFU data, the $\rm \lambda_R$ is given by
\begin{equation} \label{equ:lam}
\lambda_R \equiv \frac {\langle R|V|\rangle}{\langle R\sqrt{V^2+\sigma^2}\rangle}
\end{equation}
where  $R$, $V$ and $\sigma$ are  the circular radius,
stellar velocity and stellar velocity  dispersion, respectively. The measurement is flux weighted
and  summed  within  the effective  radius.  A  large  $\rm  \lambda_{Re}$
corresponds  to  a  fast  rotator while  a  small  $\rm  \lambda_{Re}$
indicates a slow rotator. By comparing the $\rm \lambda_{Re}$ with the
apparent  ellipticity   $\rm  \epsilon$,  \cite{Ems:11}   proposed  an
empirical diagnostic to distinguish fast and slow rotators. We measured the $\rm \lambda_{Re}$ of this galaxy based on
the MaNGA data analysis pipeline (DAP; Westfall et al. in preparation) maps.

{\bf  KINEMETRY:}  In  order  to  quantify whether  the galaxy  can  be
classified  as  a   regular   rotator  or   a  non-regular   rotator,
\cite{Kra:06}  proposed KINEMETRY\footnote{The  IDL KINEMETRY  routine
  can  be  found  on http://davor.krajnovic.org/idl.}  with  truncated
Fourier expansion to fit the velocity  fields of a galaxy with the IFU
data. Here we employed KINEMETRY to investigate the different kinematic properties varied with radius.

\section{Result}
\label{sect:analysis}

\subsection{The global galaxy properties}

The  galaxy  has  a  total   stellar  mass  of  10$^{10}$  $M_{\odot}$
\citep{Sal:07}.  Fig. 1 (a) presents the optical false-color image of the
galaxy,  which clearly  indicates that  it  is an  ETG with no sign of  spiral
arms. The central  region of the galaxy is bluer  than the outer part.

The false-color image also shows apparent extra nuclear light relative
to the outer  part of the galaxy.  To confirm  this quantitatively, we
obtain the 1-D radial profiles and fit them with S\'ersic profiles
\citep{Ser:68}. As illustrated  in Fig.  2, the  outer part of
the galaxy has a S\'ersic index of around 2.3-2.5, which is consistent with
an ETG morphology.   The inner part shows  apparent extra light
above  the  inward extrapolation  of  the  outer S\'ersic  profile.  As mentioned previously the
presence of such  extra nuclear light is typical for low-mass ETGs
\citep{Kor:99, Kor:09,  Kra:13b}.

The IFU  observations further allow  us to investigate the integrated
kinematic properties of the galaxy.  Fig. 3(a)  shows the stellar  velocity map.
To evaluate globally whether it is a slow or fast rotator, we measure
$\lambda_{R}$ within 1.0Re \citep{Ems:07,  Cap:16} and compare it with
the apparent  ellipticity $\epsilon$.   As shown in  Fig. 3(b),  the galaxy
lies above  the dividing line  in spite of  measurement uncertainties,
indicating that it  is a fast-rotator, again  consistent with low-mass
ETGs.

\subsection{ An inner rotating disk with ongoing star formation}

The inner  and outer  regions of the galaxy show  distinct kinematic  properties. As
shown in Fig. 3 (a), the nuclear region containing the extra
light is  clearly a rotating disk  while the outer part  is dispersion
dominated.   This is  further demonstrated  in the  $\lambda_{\rm Re}$
vs. $\epsilon$ plane as shown in Fig. 3 (b): the inner component is above
the dividing line between slow and fast rotators, while the outer component
is below the  line. The galaxy thus  belongs to the type  of ETGs with
kinematically   decoupled  cores   as   found   in  previous   surveys
\citep{Kor:09,  Kra:11, Che:12}. As it  is impossible to  produce such
cores through any internal process,  the presence of the kinematically
decoupled core is proof of galaxy merging.  We further used the
result of the  KINEMETRY to quantify the kinematic  properties of both
parts.  As shown  in Fig.  4,  in the  inner  region the  positon angle (P.A.) of  the
best-fitted  ellipse is  almost  constant, and  the  velocity of  each
ellipse  dominates  over  the  dispersion so  that  $\rm  k_1/k_5$  is
small. Furthermore, from the rotation curve in the inner region, we confirm that this can't be a bar since it doesn't exhibit a rigid-body rotation \citep{Kor:82}. In contrast,  ellipses orient randomly in the outer region and
the  velocity  drops to  almost  zero  with $\rm  k_1/k_5$  increasing
rapidly, which indicate the  dominance of dispersion. We also
used the  methodology   in  \cite{Tab:17}   to  make   a  kinematic
decomposition of the galaxy, and confirmed that the inner part is disk
like while the outer part is bulge like.

Fig. 1 (b) is the $g$ band image from BASS (Beijing-Arizona Sky Survey) data release 2 \citep{Zou:17,Zou:sub} which is 0.87 mag deeper than the SDSS. It covers $227 \times 227$ arcsec$^2$ FOV, about twenty times bigger than the coverage of the Fig. 1 (a). There are some faint tidal features around the host galaxy. After smoothing the image following the method in \cite{Mis:11}, the tidal features are more obvious, as exhibited in Fig. 1 (c). These tidal features further indicate the recent or ongoing merger events \citep{,Mar:10,Mis:11}. 

As shown in Fig. 3 (a), we  defined the inner region with a major axis
of 4.33'' (1.54 kpc) and an axis ratio of 0.5. Following the methodology described in \cite{Sal:07} and \cite{Che:10}, we derived the stellar M/L of inner region by comparing the five bands colors with the grid of CB08 \citep{CB08} model galaxies colors. Then we obtained the stellar mass of the
inner region is 10$^{9.5}$ M$_{\odot}$,  accounting for 30\% of the total
stellar mass of the galaxy.   This large fraction suggests the merger event which 
formed the disk is likely to have been a major merger.

In contrast to  the kinematically decoupled cores  in other elliptical
galaxies, the one in SDSS J142055.01+400715.7 harbors ongoing star formation, as
indicated by the spatially resolved BPT diagram \citep{Bal:81, Kew:01, Kau:03}. As shown in Fig.  5, the
rotation-dominated disk is filled with star forming regions, while the
outer region has LIER-like \citep{Bel:16} emission likely excited by  old stellar
populations. The star formation rate is calculated from the H-$\alpha$
emission, accounting for  dust  extinction which we calculated using the Balmer
decrement    along    with    the   Calzetti's    attenuation    curve
\citep{Cal:00}. As  shown in  Fig. 5  (c), the  inner disk  is located
within the main sequence of star-forming galaxies, indicating that its
level of  star formation  is more  like that of a spiral  galaxy than a 
starburst.   This  presence of  ongoing  star  formation offers  direct
evidence that the merging responsible for this complex galaxy structure must be gas rich.

\section{Discussion}
\label{sect:discu}

While  it has  long been  suspected that  low mass  ETGs form  through
gas-rich merging/accretion, it is been difficult to associate them directly
with  gas-rich  processes because the vast majority of ETG remnants have long since ceased the phase of
 ongoing star formation indicative of these processes.
We have found a galaxy that appears to  be in the final  phase of gas-rich
merging,  the result being  a typical low-mass ETG.  While globally  it is a
fast-rotating  ETG with  extra nuclear  light, the  distinct kinematic
features of the inner and outer parts and the tidal features surrounding the host galaxy suggest galaxy merging as
the  origin of  this galaxy,  and the  ongoing star  formation in  the
nuclear region further indicates that the merging is gas rich. The cold
gas mass can be roughly estimated  with the star formation law that is
the relationship between the star  formation rates and gas masses \citep{Ken:98}.  We
first estimate the oxygen  abundance to  be Solar  so that  the star
formation law of spiral galaxies  can be used \citep{Shi:14}. Based on
the equation by \citet{Shi:18}, the derived cold gas mass is about 10$^{8.13}$
M$_{\odot}$. The gas  to stellar mass ratio is thus  small, only about
1\%. This may be  because the object is already at  the final stage of
merging  so that the  majority of the  gas has been  expelled or
consumed.

The galaxy also offers direct evidence that the extra nuclear light is
contributed  by  star  formation  activity in  the  nuclear  regions
\citep[e.g.][]{Kor:99}. The spatial extent  of the extra nuclear light
as seen in the 1-D radial profile is  more or less the same as that of
star-formation regions, as seen in the  spatially-resolved BPT diagram
(Fig. 5). The extra nuclear light  covers a region with radius around
0.7-1.0 kpc  in the optical  band (Fig. 2 and Fig. 6 (b)), which is  much larger than observed in other ETGs (a  few$\times$0.1 kpc) (Kormendy et al. 2009).
This is  likely caused by the  young age of the  stellar population in
the extra  light so that  the mass-to-light  ratio is smaller  for the
inner part.

Several studies have found that some ETGs exhibit ongoing star formation. These ETGs often contain gas and stellar misalignment, indicating fresh gas accretion from outside \citep{Ily:14,Sil:14}. However, they are different from the galaxy in the current work. Those galaxies are already ETGs and acquire a minor amount of external gas to trigger moderate star formation, sometimes referred to as rejuvenation. On the other hand, our galaxy lies in the star forming main sequence, indicating gas-rich merging. In Fig. 6 we plot $\rm H\delta-Dn4000$ diagram \citep{Bal:99} for different radii. The best fitting rings are extracted from the ``ellipse'' task in IRAF. There is a monotonic decrease outward, indicating an outside-in quenching scenario. This could be evidence of the final phase of gas-rich merging as it produces a low-mass ETG.

\section{Conclusion}
\label{sect:conclu}

We report  a discovery  of an early-type galaxy in MaNGA that  offers direct
evidence for gas-rich merging as  the origin of low-mass ETGs. The
integrated  properties of  the galaxy  are consistent  with a  typical
low-mass ETG with $M_{*}$=10$^{10}$  M$_{\odot}$, no spiral arms, fast
rotating kinematics and an outer  S\'ersic profile index of 2.1-2.4. However, the outer
and inner  parts show distinct dispersion dominated
and rotation  dominated kinematics, respectively. And the broad-band image shows some tidal features surrounding the galaxy.  These all suggest recent galaxy
merging.  Furthermore  the inner disk harbors  on-going star formation
as indicated by the BPT diagram,  which suggests that the merging is gas
rich.  The  type of the  galaxy is
  rare but it may be a demonstration of the role that gas-rich merging plays in making a
  low-mass ETG.\\

{\noindent \bf Acknowledgements}

We thank the referee for a detailed report which helped significantly improve the presentation of our work. S.L. and Y.S. acknowledge support from the National Key R\&D Program of China (No. 2018YFA0404502), the National Natural Science Foundation of China (NSFC grants 11733002 and 11773013), the Excellent Youth Foundation of the Jiangsu Scientific Committee (BK20150014), and National Key R\&D Program of China (No. 2017YFA0402704). Y.C. acknowledges support from National Natural Science Foundation of China (NSFC grants 11573013). D.B. is supported by grant RScF 14-50-00043. Funding for the  Sloan Digital Sky Survey IV has  been provided by the
Alfred P.  Sloan Foundation, the  U.S. Department of Energy  Office of
Science,  and the  Participating Institutions.  SDSS- IV  acknowledges
support and  resources from the Center  for High-Performance Computing
at the University of Utah. The SDSS web site is www.sdss.org. SDSS-IV is  managed by the  Astrophysical Research Consortium  for the
Participating  Institutions of  the SDSS  Collaboration including  the
Brazilian Participation  Group, the Carnegie Institution  for Science,
Carnegie  Mellon  University,  the Chilean  Participation  Group,  the
French   Participation    Group,   Harvard-Smithsonian    Center   for
Astrophysics,  Instituto de  Astrof\'{i}sica  de  Canarias, The  Johns
Hopkins University, Kavli Institute for the Physics and Mathematics of
the Universe (IPMU) / University  of Tokyo, Lawrence Berkeley National
Laboratory,  Leibniz  Institut   f\"{u}r  Astrophysik  Potsdam  (AIP),
Max-Planck-Institut    f\"{u}r     Astronomie    (MPIA    Heidelberg),
Max-Planck-Institut     f\"{u}r     Astrophysik    (MPA     Garching),
Max-Planck-Institut f\"{u}r Extraterrestrische  Physik (MPE), National
Astronomical Observatory  of China,  New Mexico State  University, New
York University, University of Notre Dame, Observat\'{o}rio Nacional /
MCTI,  The  Ohio  State  University,  Pennsylvania  State  University,
Shanghai Astronomical Observatory, United Kingdom Participation Group,
Universidad  Nacional   Aut\'{o}noma  de  M\'{e}xico,   University  of
Arizona,  University  of  Colorado   Boulder,  University  of  Oxford,
University of Portsmouth, University  of Utah, University of Virginia,
University   of  Washington,   University  of   Wisconsin,  Vanderbilt
University, and Yale University.

\begin{figure*}
  \resizebox{15cm}{!}{\includegraphics{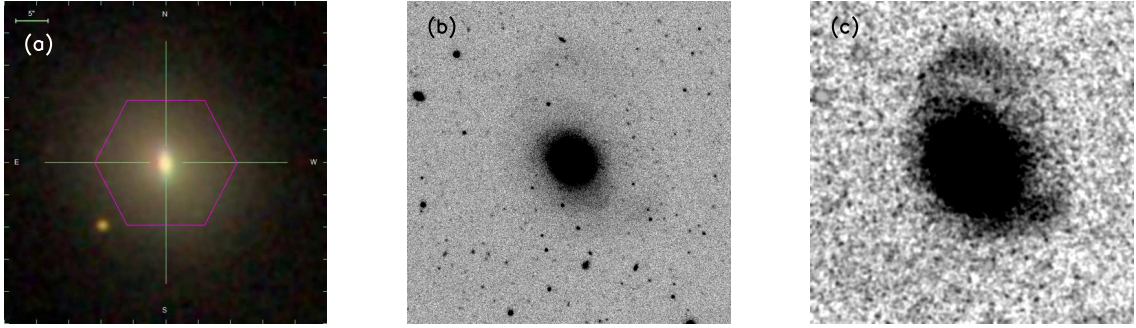}}
  \caption{{\bf a:} The false color image of SDSS J142055.01+400715.7, the hexagon shows the coverage of MaNGA bundle; {\bf b:} The BASS $g$ band image of this galaxy covering a field of $227 \times 227$ arcsec$^2$; {\bf c:} The smoothed image of Fig. 1 (b) adopting the procedure from \citet{Mis:11}.}
\end{figure*}

\begin{figure*}
  \resizebox{15cm}{!}{\includegraphics{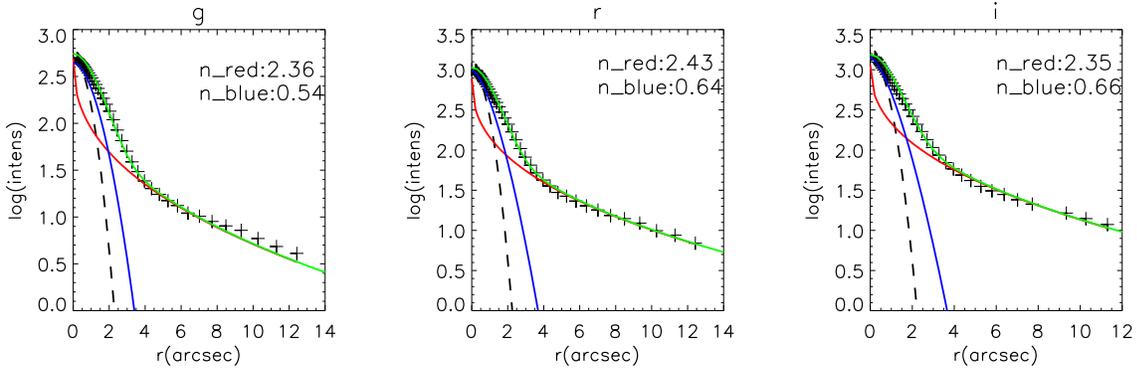}}
  \caption{The radial profiles of the object in $g$, $r$ and $i$ bands. The symbols are fitting points from ellipse task in IRAF. The dashed lines are simplified Gaussian PSF and the blue and red lines are two intrinsic S\'ersic profiles. The S\'ersic indices are listed in each panel. The green lines are the sum of these two S\'ersic profiles after convolving with the PSF.}
\end{figure*}

\begin{figure*}
  \resizebox{7cm}{!}{\includegraphics{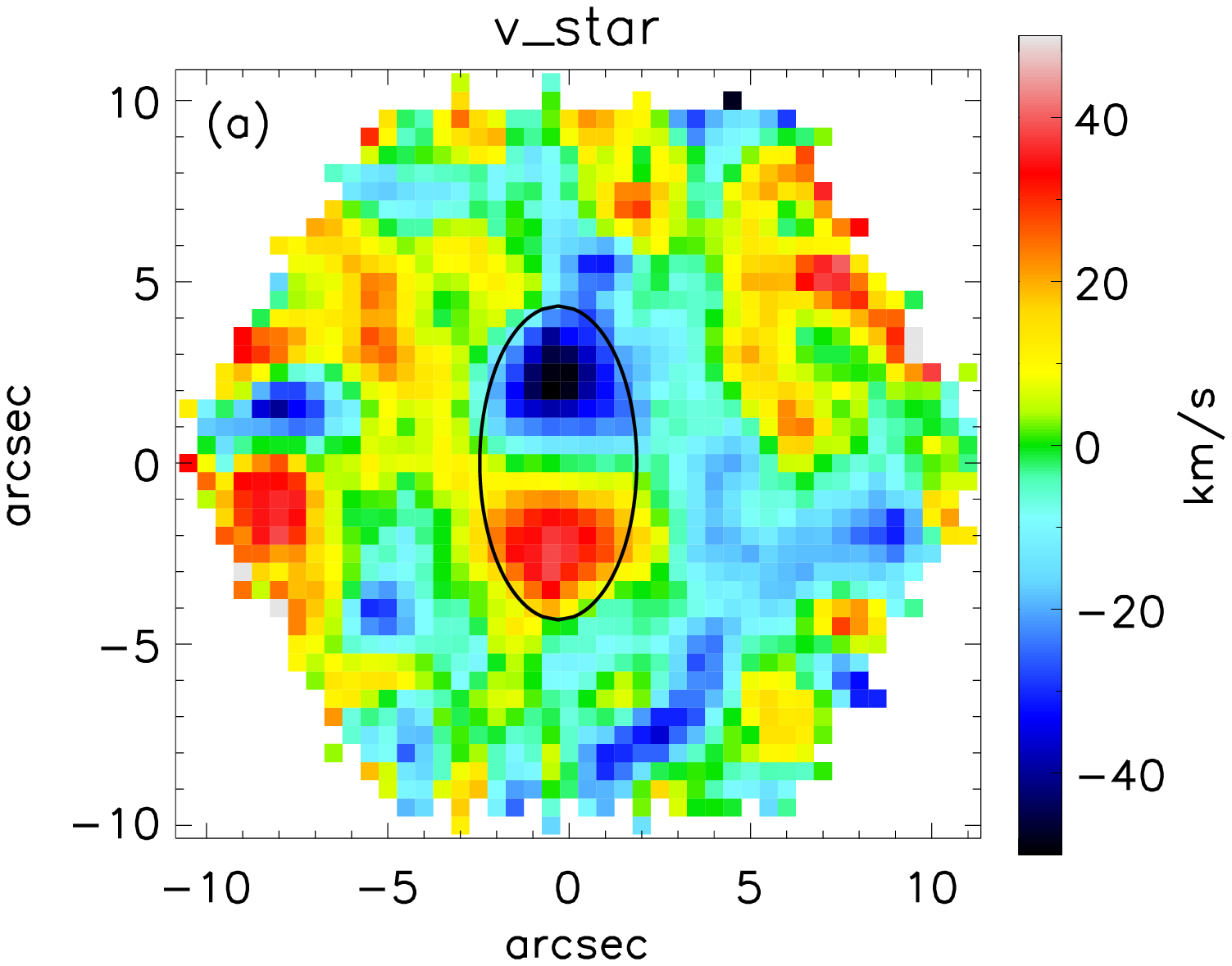}}
  \resizebox{7cm}{!}{\includegraphics{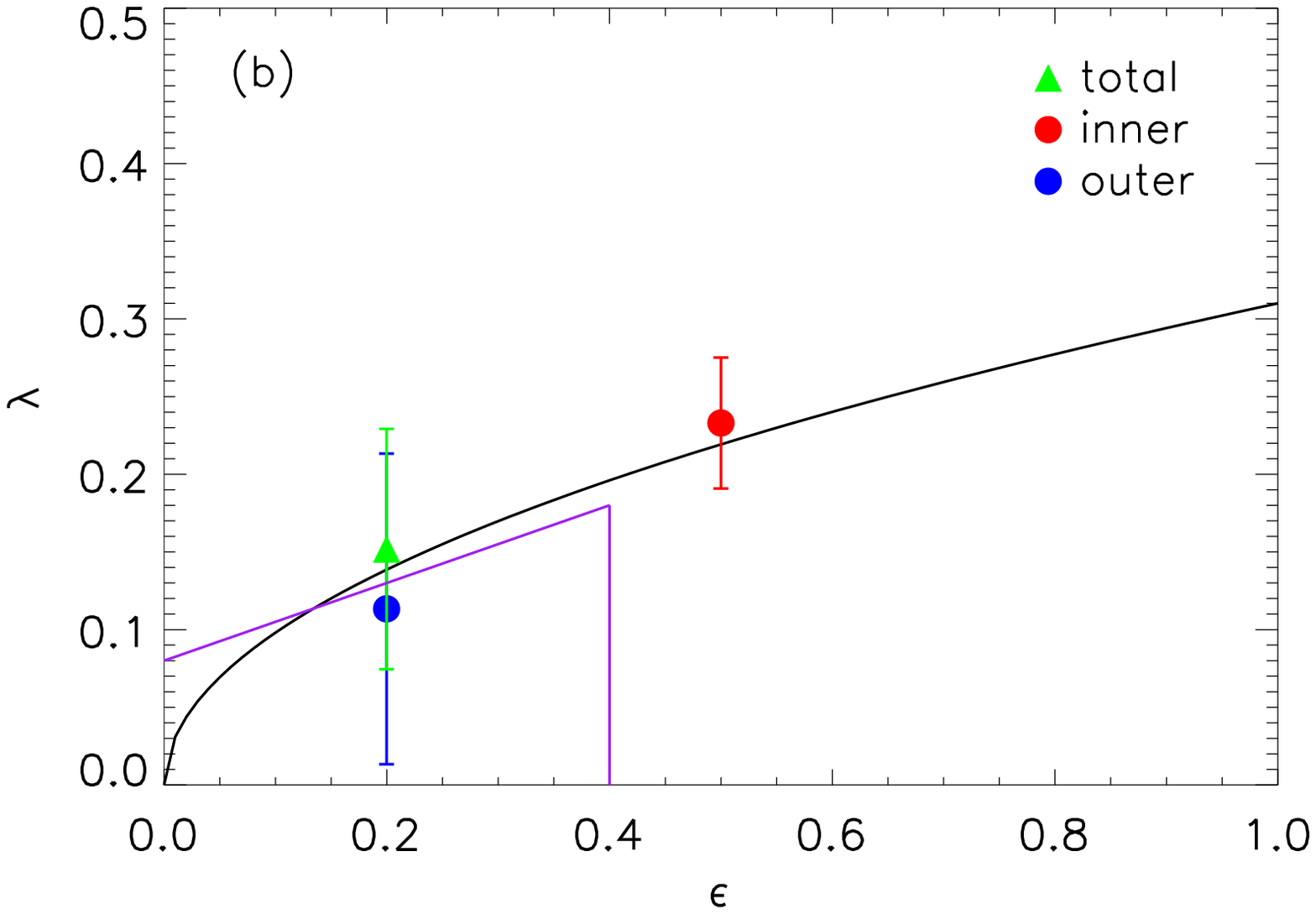}}
  \caption{{\bf a:} The stellar velocity map of the object. Ellipse encloses the inner region we chose manually; {\bf b:} The location of the object on the  $\lambda_{\rm Re}$ vs. $\epsilon$ plane. The solid black line is the the critical line to distinguish fast and slow rotators proposed by \citet{Ems:11}, which is $\rm \lambda_{Re}=0.31\times\sqrt{\epsilon_e}$. The purple line is the fast/slow rotator division proposed in \citet{Cap:16} to reduce the risk of missing very round regular rotators.}
\end{figure*}

\begin{figure*}
  \resizebox{5cm}{!}{\includegraphics{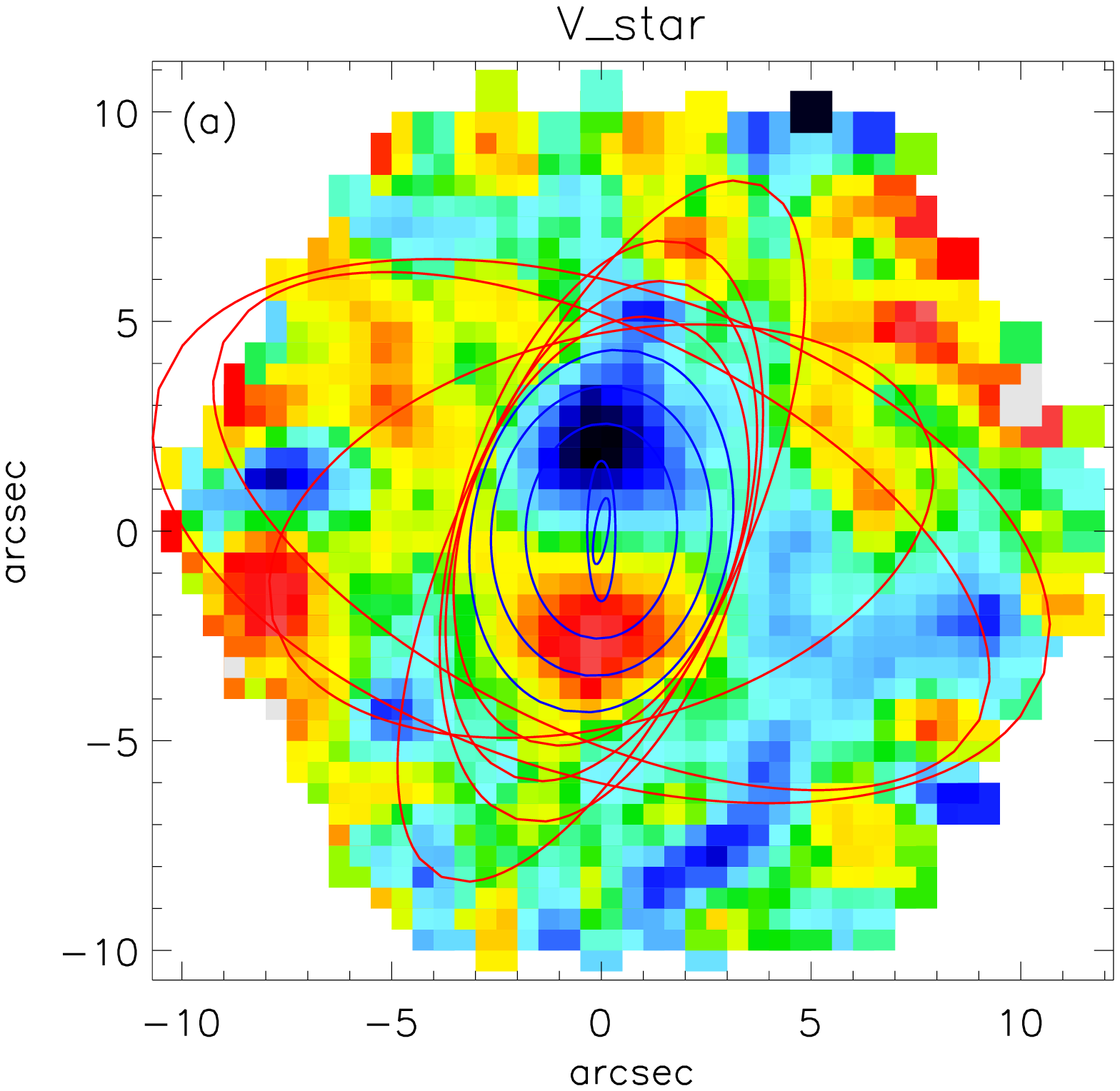}}
  \resizebox{5cm}{!}{\includegraphics{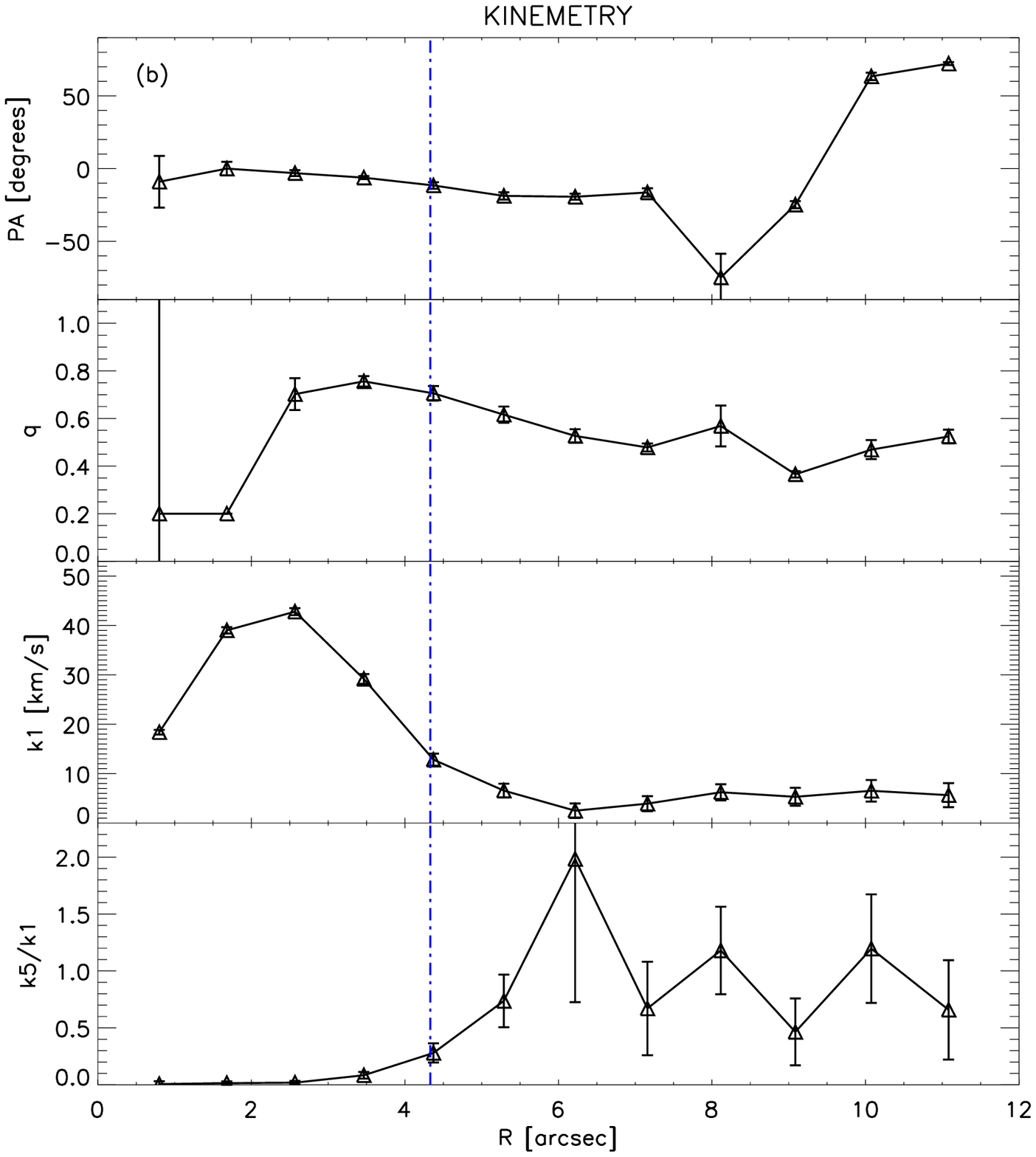}}
  \caption{{\bf a:} The stellar velocity map of the object overlaid with the best-fitted
  kinematic ellipses. The blue ellipses correspond to the radius smaller than inner part chosen from fig. 3 (a). {\bf b:} The fitting result of the KINEMETRY. The blue line is the boundary of the inner region.}
\end{figure*}

\begin{figure*}
  \resizebox{5.5cm}{!}{\includegraphics{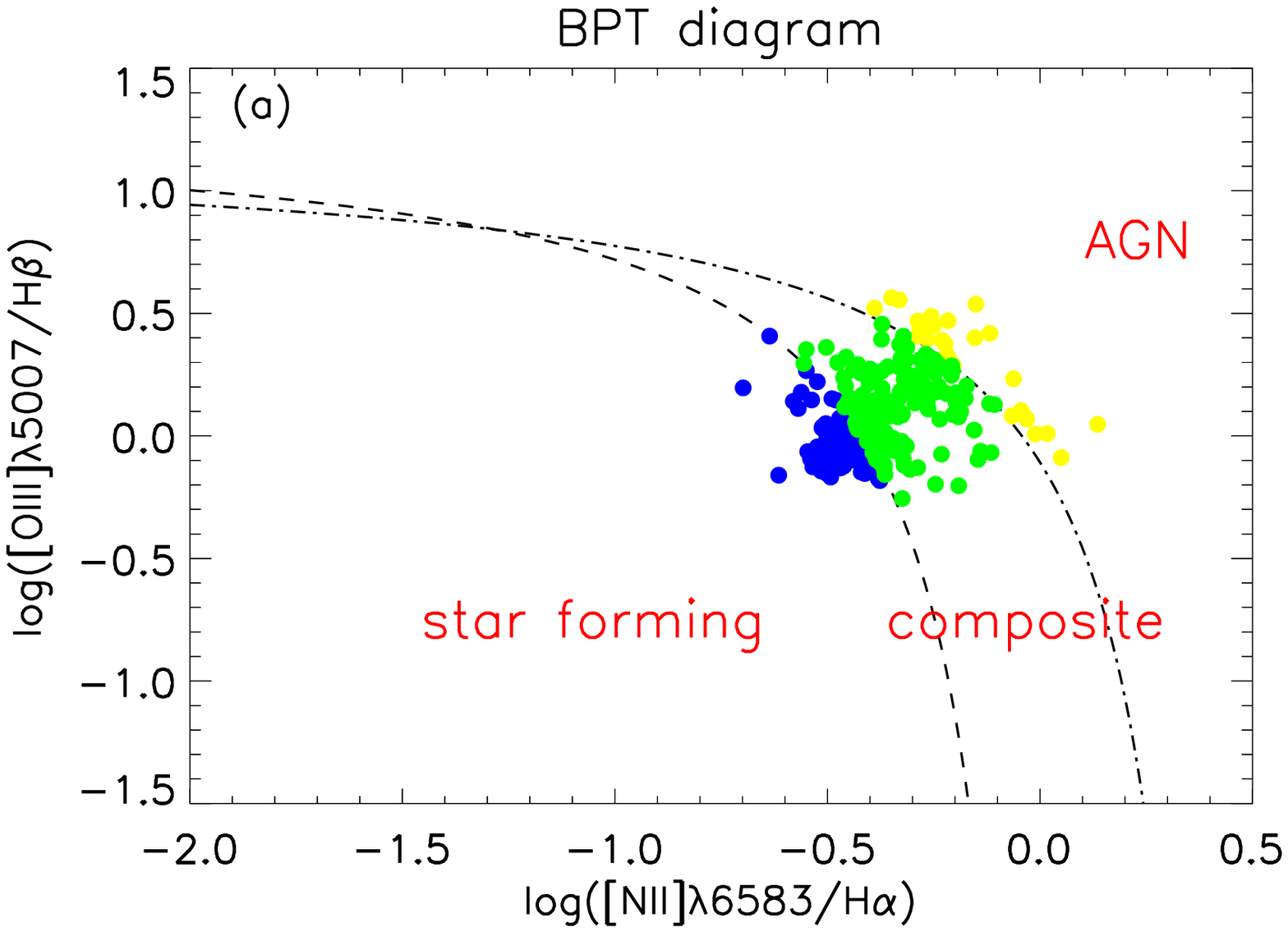}}
  \resizebox{3.2cm}{!}{\includegraphics{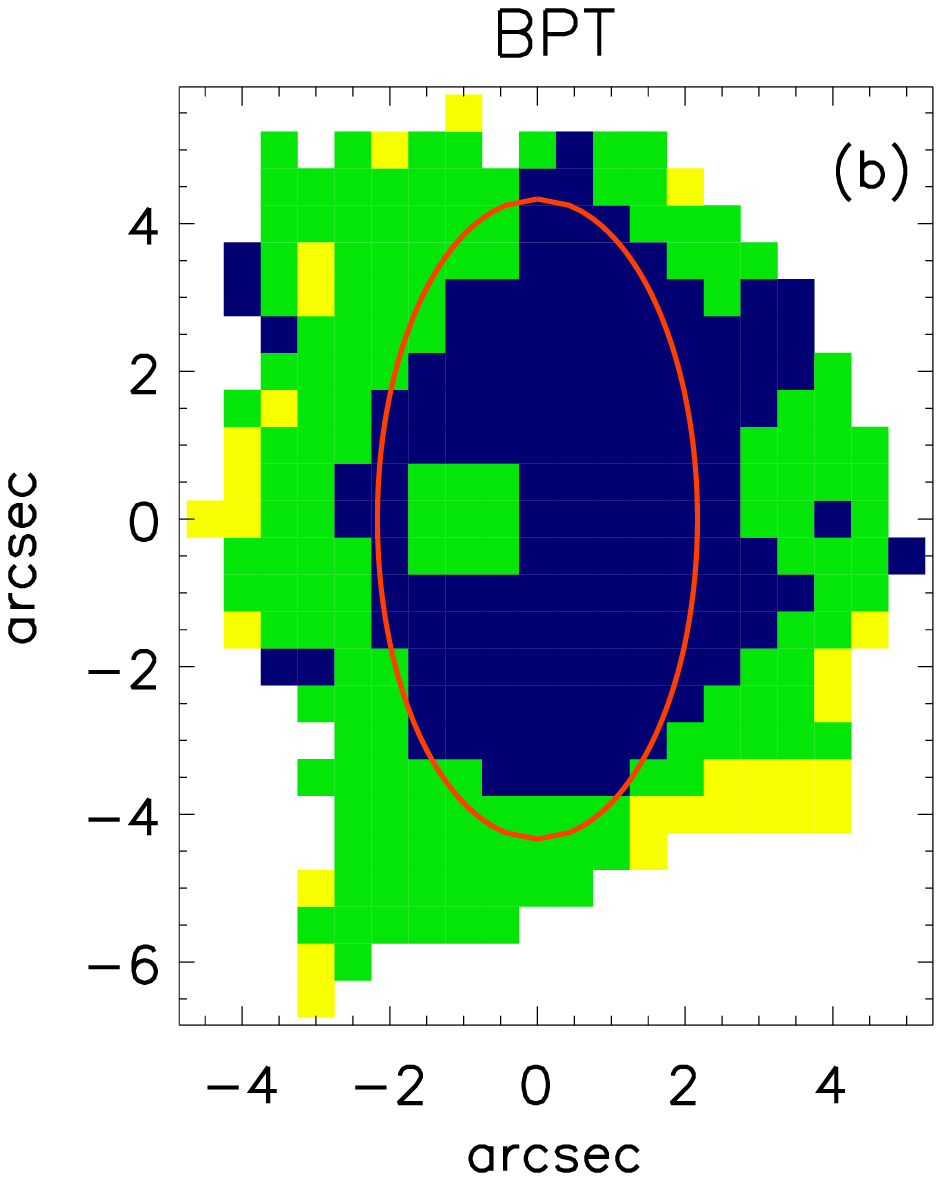}}
  \resizebox{3.8cm}{!}{\includegraphics{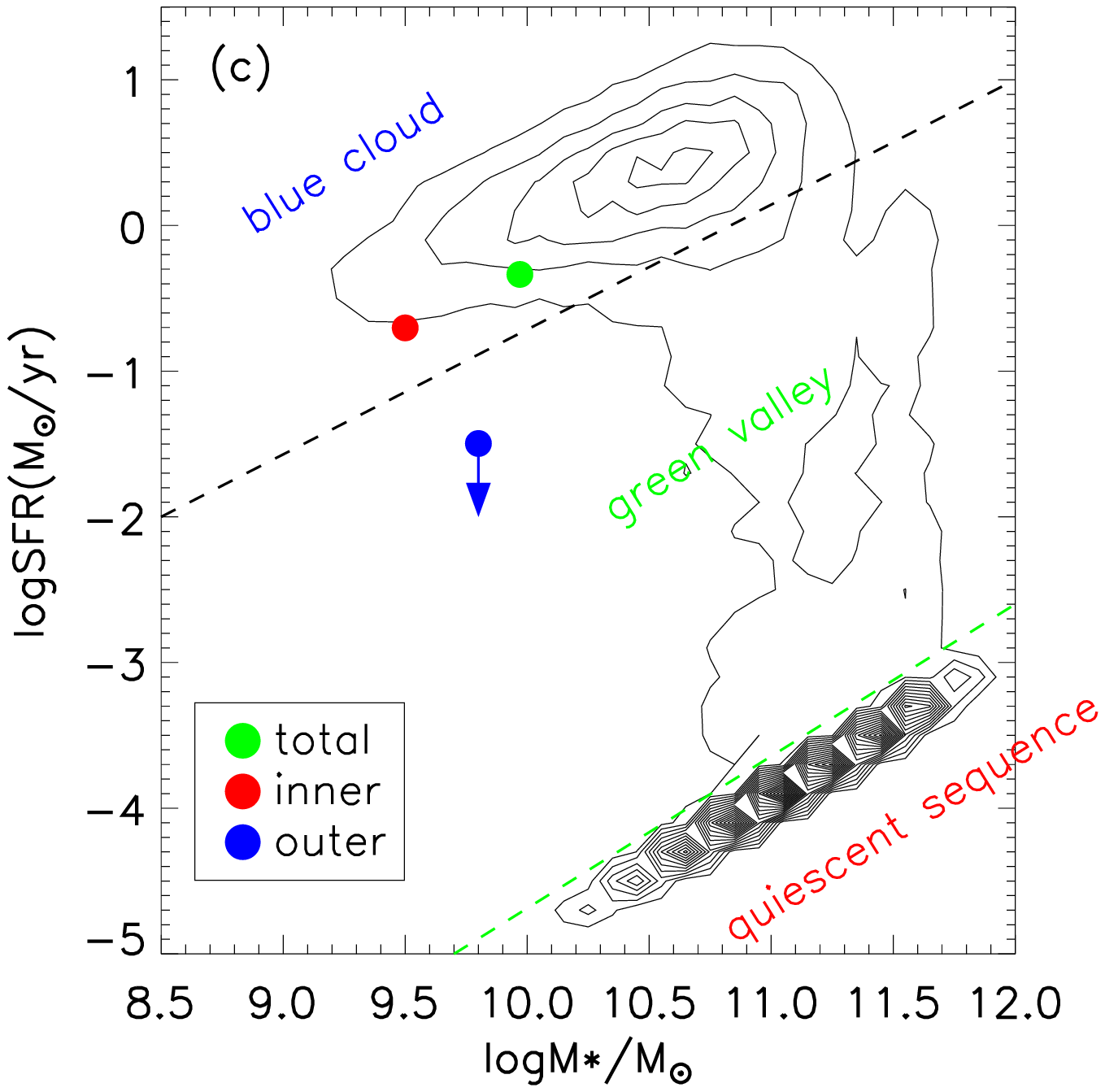}}
  \caption{{\bf a:} The BPT diagram for individual pixels. the dash-dot line is the theoretical ``maximum starburst line'' proposed by \citet{Kew:01}. The dashed line is proposed by \citet{Kau:03} to rule out possible composite galaxies whose spectra contain significant contributions from both AGN and star formation. {\bf b:} The spatially resolved BPT map. The colors correspond to those in Fig. 5 (a). The The red ellipse is the boundary of the inner part.  {\bf c:} The location of the object on the SFR vs. stellar mass
  plane. The contours show 100000 samples from SDSS+WISE MAGPHYS CATALOG \citep{Cha:15}. }
\end{figure*}

\begin{figure*}
  \resizebox{5cm}{!}{\includegraphics{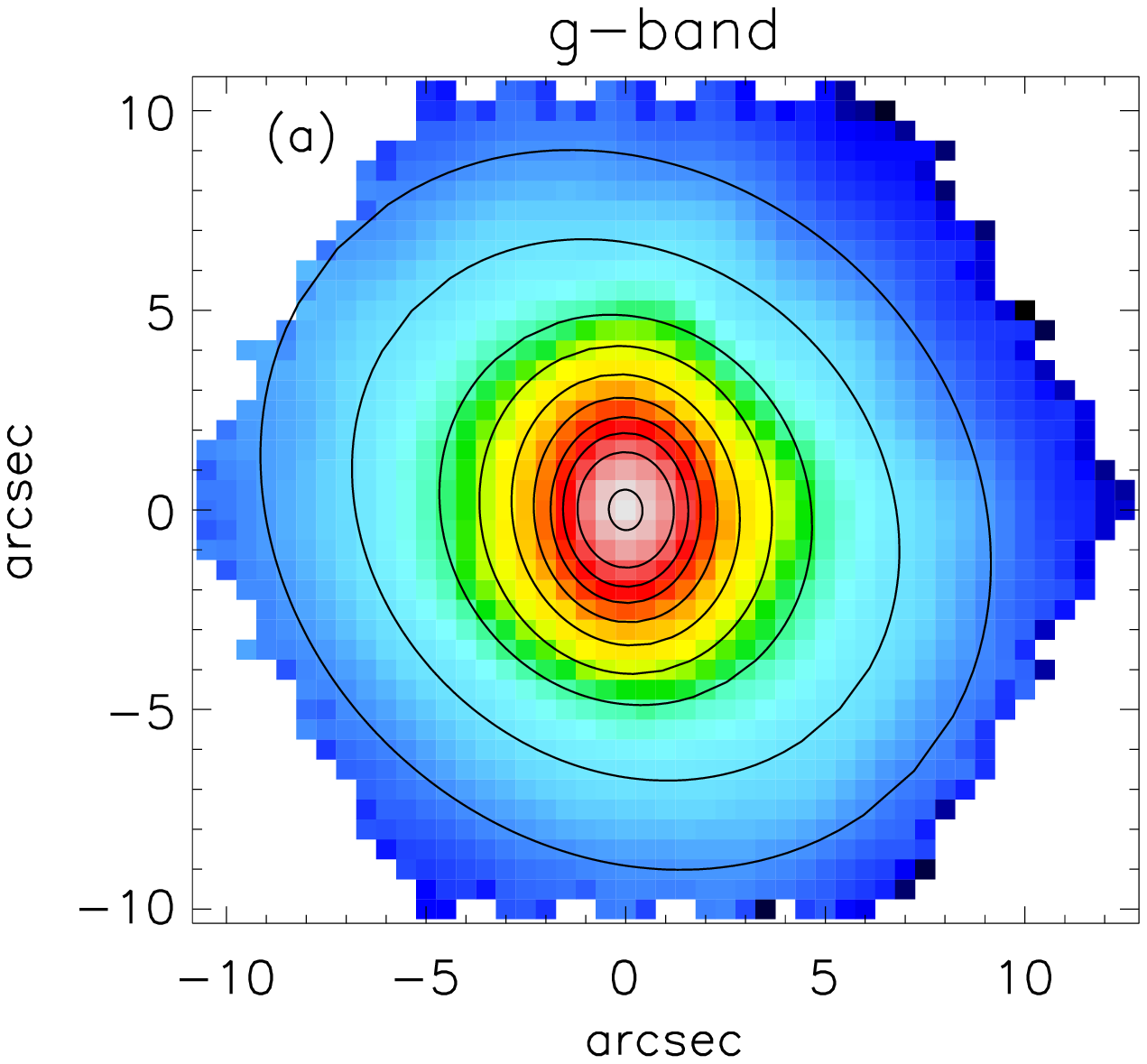}}
  \resizebox{7cm}{!}{\includegraphics{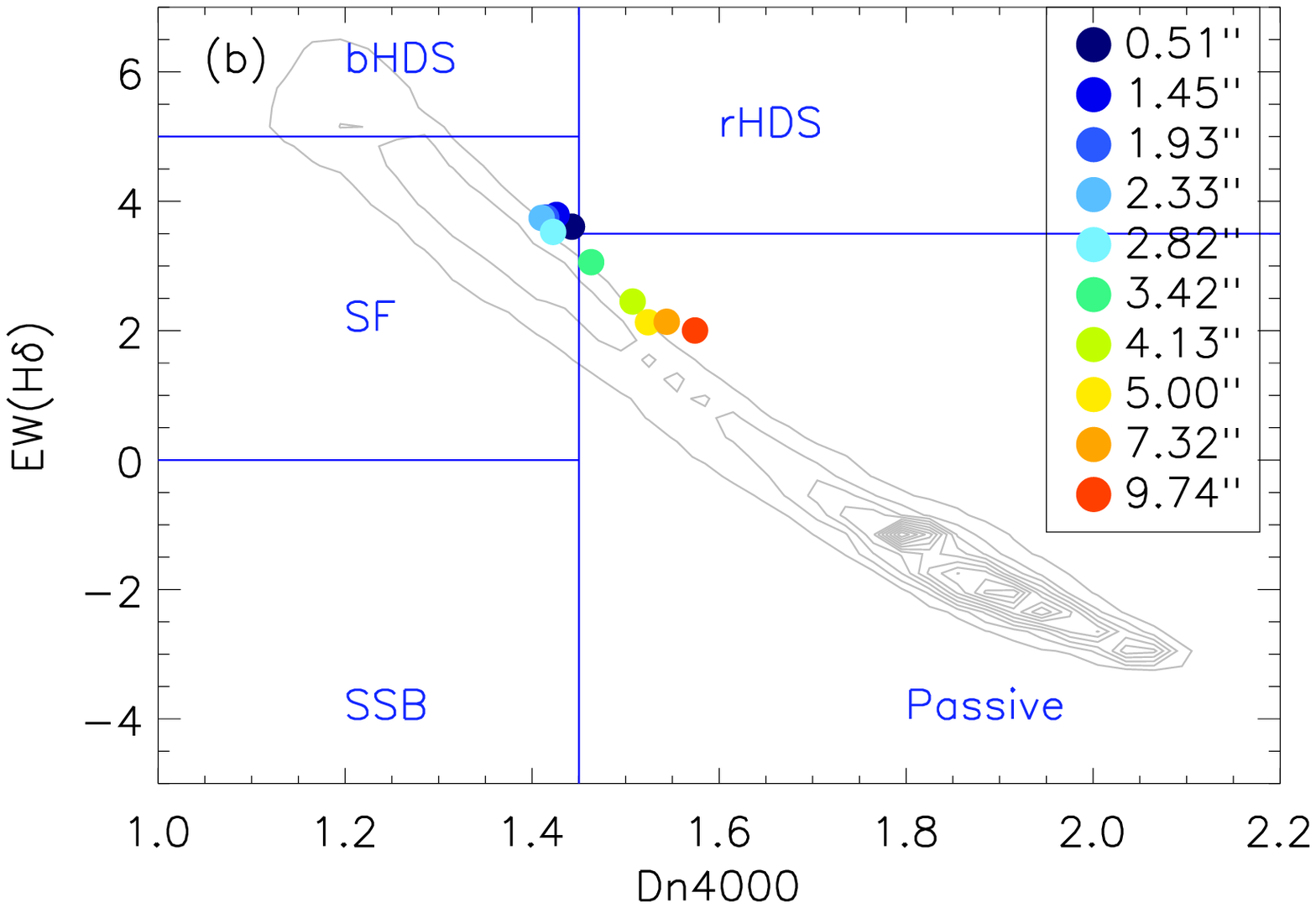}}
  \caption{{\bf a:} The g band flux map in logarithmic scale. Ten ellipse rings are extracted from ``Ellipse'' task in IRAF. {\bf b:} The $\rm H\delta-Dn4000$ diagram of 10 rings in Fig. 6 (a). The contour shows 100000 SDSS DR7 galaxies \citep{Bal:99,Wor:97}. The classifications are from \citet{Bal:99}.}
\end{figure*}

\bsp

\label{lastpage}
\end{document}